\documentclass[aps,prl,twocolumn,groupedaddress]{revtex4-1}

\usepackage[utf8]{inputenc}
\usepackage[T1]{fontenc}
\usepackage{lmodern}
\usepackage{amsmath,amsfonts,amssymb}
\usepackage{graphicx}

\usepackage{hyperref}
\usepackage[usenames,dvipsnames]{xcolor}
\hypersetup{colorlinks=true, linkcolor=blue!50!black, urlcolor=blue!50!black, citecolor=blue!50!black}
\usepackage[all]{hypcap}

\newcommand{\img}{\mathsf{i}}
\newcommand\diff{\mathrm{d}}
\renewcommand{\vec}[1]{\mathbf{#1}}

\newcommand{\subref}[2]{\hyperref[#1]{#2}}

\usepackage[normalem]{ulem}

\begin{document}

\title{Time-Dependent Fluctuations and Superdiffusivity in the Driven Lattice Lorentz Gas}



\author{Sebastian Leitmann}
\affiliation{Institut f\"ur Theoretische Physik, Universit\"at Innsbruck, Technikerstra{\ss}e~21A, A-6020 Innsbruck, Austria}
\author{Thomas Franosch}
\affiliation{Institut f\"ur Theoretische Physik, Universit\"at Innsbruck, Technikerstra{\ss}e~21A, A-6020 Innsbruck, Austria}
\email[]{thomas.franosch@uibk.ac.at}

\date{\today}

\begin{abstract}
We consider a tracer particle on a lattice in the presence of immobile obstacles. Starting from
equilibrium, a force pulling on the particle is switched on, driving the system to a new stationary state. We solve for
the complete transient dynamics of the fluctuations of the tracer position along the direction of the force. The analytic
result, exact in first order of the obstacle density and for arbitrarily strong driving, is compared to stochastic
simulations. Upon strong driving, the fluctuations grow superdiffusively for intermediate times; however, they 
always become diffusive in the stationary state. The diffusion constant is
nonanalytic for small driving and is enhanced by orders of magnitude by increasing the force.
\end{abstract}


\maketitle




The material properties of complex fluids such as colloidal dispersions~\cite{Hunter:RPP_75:2012,Wilson:JPCB_113:2009},
solutions of biopolymers~\cite{Amblard:PRL_77:1996,Koenderink:PRL_96:2006}, or
biomaterials~\cite{Bausch:BioPhys_76:1999,Mizuno:SCI_315:2007} can be probed by pulling a mesoscopic tracer particle through the medium.
In linear response by the fluctuation-dissipation theorem it is sufficient to monitor the force-free thermally agitated
motion of the tracer which is the principle of passive
microrheology~\cite{Mason:PRL_74:1995,Waigh:RPP_68:2005,Squires:ARFM_42:2010}. Then by a generalized Stokes-Einstein
relation the dynamic mobility is connected to the linear macroscopic frequency-dependent viscosity or elastic modulus. In
contrast, in \textit{active} microrheology the particle is manipulated by optical or magnetic tweezers and pulled through the
environment in principle by arbitrarily strong forces~\cite{Squires:LM_24:2008,Wilson:PCCP_13:2011,Puertas:JPCM_26:2014}. Here, the
system is intrinsically strongly out of equilibrium and a plethora of
new phenomena have been found experimentally and in simulations, such as
force thinning~\cite{Habdas:EPL_67:2004,Carpen:JRHEO_49:2005,Sriram:PoF_22:2010}, (transient) superdiffusive
behavior, and enhanced diffusivites~\cite{Winter:PRL_108:2012,Winter:JCP_138:2013}.

To make theoretical progress in the nonlinear regime, generic models have been investigated that focus on the mutual
exclusion originating from the strong repulsive interaction between the tracer and its environment as the most important
ingredient.
The underlying dynamics of the tracer
is usually modeled as a random walk on a lattice or Brownian motion in continuum, while the surroundings range from 
dilute and immobile obstacles to dynamic and crowded environments. For lattice systems, progress and even exact results
have been achieved~\cite{Jack:PRE_78:2008,Leitmann:PRL_111:2013,Basu:JPAMT_47:2014,Baiesi:PRE_92:2015,Illien:PRL_113:2014,
Benichou:PRL_113:2014,Illien:JSMTE:2015,Benichou:PRE_93:2016}, predicting \textit{inter alia} anomalous diffusion~\cite{Illien:PRL_111:2013}
and superdiffusive behavior~\cite{Benichou:PRL_111:2013} in confined systems.
In continuum, the framework of mode-coupling theory of the glass
transition~\cite{Gazuz:PRL_102:2009,Gnann:SM_7:2011,Gnann:PRE_86:2012,Harrer:JPCM_24:2012,Gazuz:PRE_87:2013,Wang:PRE_89:2014,Gruber:PRE_94:2016}, Langevin
equations~\cite{Demery:NJP_16:2014,Demery:PRE_91:2015}, kinetic theory~\cite{Wang:PRE_93:2016}, and continuous-time random
walks~\cite{Schroer:PRL_110:2013,Schroer:JCP_138:2013,Burioni:CTP_62:2014} successfully describe
certain phenomena emerging in the nonlinear regime. Exact results in the stationary state in first order of the bath
particles have been obtained for active microrheology in suspensions of hard spheres performing
Brownian motion~\cite{Squires:PoF_17:2005,Khair:JFM_557:2006,Zia:JFM_658:2010,Swan:PoF_25:2013,Hoh:JFM_795:2016}; yet, an evaluation of
the transient dynamics and the approach to the steady state has remained a challenge. 

Here we rely on a lattice model for a driven tracer in a crowded environment to investigate the growth
of the fluctuations as time progresses. The crowding is incorporated in the model by introducing hard and immobile
obstacles randomly distributed over the lattice. A force is switched on at a certain instant of time 
such that the tracer performs a biased obstructed diffusion through the system. To first order in the obstacle density
the moment-generating function for the displacements can be determined in principle exactly; so far only the time-dependent
velocity response has been elaborated~\cite{Leitmann:PRL_111:2013}. In equilibrium, the fluctuations are also known for low
obstacle densities via the mean-square displacement~\cite{Nieuwenhuizen:PRL_57:1986,Nieuwenhuizen:JPAMG_20:1987,Ernst:JPAMG_20:1987}. 
Within this model, we consider the 
fluctuations along the direction of the force and provide for the first time a complete time-dependent analytic solution
for a generic strongly interacting system driven far from equilibrium.

\begin{figure*}
\includegraphics[scale=0.68]{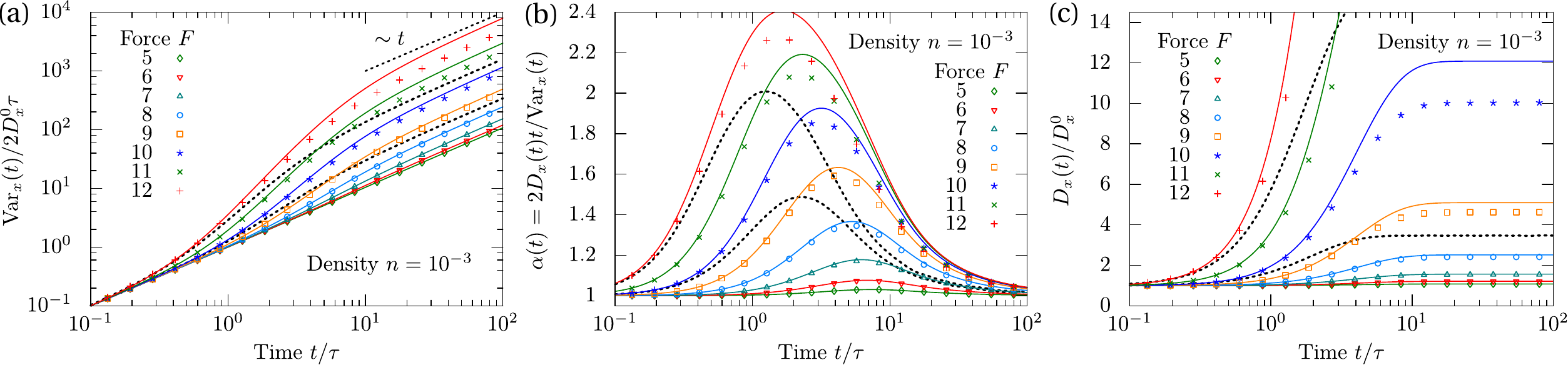}
\caption{ \label{fig:varx_d1e-3} \label{fig:diffcoeffx_d1e-3} \label{fig:subsuperdiffx_d1e-3} 
Fluctuations of the tracer along the force characterized by (a) the variance $\text{Var}_x(t)$, (b) the local
exponent $\alpha(t)$, and (c) the time-dependent diffusion coefficient $D_x(t)$. In all three panels, force increases
from bottom to top. Solid lines correspond to the analytic solution and symbols represent results from computer
simulations. The black dashed lines represent simulation results for $F=10$ and $12$ with mobile
obstacles at the same density performing a symmetric random walk with mean waiting time $\tau$.
}
\end{figure*}

\textit{Model.---} 
We consider a tracer particle performing a random walk with successive nearest-neighbor jumps $\mathcal{N} = \{\pm
a\vec{e}_x, \pm a\vec{e}_y\}$ on a square lattice with lattice spacing $a$. The lattice consists of free sites
accessible to the tracer, as well as sites with randomly placed immobile hard obstacles of density $n$.  If the tracer
attempts to jump onto an obstacles site, it merely remains at its original position.

For times $t < 0$ the tracer performs a symmetric random walk and the system is in equilibrium, such that it is equally
likely to find the tracer at any accessible site. For times $t \geq 0$, we apply a force pulling the tracer along the $x$ direction of
the lattice. The dimensionless force 
$F = \text{force} \cdot a/k_\text{B} T$ introduces a bias in the nearest-neighbor
transition probabilities $W(\vec{d}\in \mathcal{N})$, and local detailed balance $W(a\vec{e}_x)/W(-a\vec{e}_x)=\exp(F)$ and
$W(a\vec{e}_y)/W(-a\vec{e}_y)=1$ along both lattice directions suggests $W(\pm a\vec{e}_x) = e^{\pm
F/2}/(e^{F/2} + e^{-F/2} + 2)$ and $W(\pm a\vec{e}_y) = 1/(e^{F/2}+ e^{-F/2} + 2)$. 
We perform computer simulations of this model and monitor the displacement along the force $\Delta x_j = x_j - x_0$ in
discrete time corresponding to the number of (attempted) jumps $j$ of the tracer particle. The trajectories are averaged over 
many initial positions and obstacle realizations and transformed to
continuous time $t$ via a Poisson transform~\cite{Haus:PR_150:1987}:
\begin{align}
\langle \Delta x(t)^m \rangle = \sum_{j = 0}^\infty \langle \Delta x_j^m \rangle \frac{(\Gamma t/\tau)^j}{j!} e^{-\Gamma
t/\tau},
\end{align}
with mean waiting time $\tau/\Gamma = 2\tau/[1+\cosh(F/2)]$ of the tracer, valid for any order $m$ of the moment.
The choice for the mean waiting time corresponds to unnormalized transition rates $(\Gamma/\tau)
W(\vec{d})$; in particular, the transition rate perpendicular to the force is independent of the driving.
Considering normalized rates as in Ref.~\cite{Leitmann:PRL_111:2013} results only in a (force-dependent) multiplicative shift of the time scale.

Here, the quantity of interest is the time-dependent fluctuations of the tracer position along the direction of the force, 
encoded in the variance of the displacement:
\begin{align} \label{eq:variance_general}
\text{Var}_x(t) = \bigl\langle[\Delta x(t) - \langle\Delta x(t)\rangle]^2\bigr\rangle = \langle \Delta x(t)^2\rangle - \langle \Delta x(t) \rangle^2.
\end{align}
For the special case without driving, $F = 0$, the lattice Lorentz gas is
recovered and an analytic solution for the time-dependent fluctuations in first order of the obstacle density was achieved
years ago~\cite{Nieuwenhuizen:PRL_57:1986,Nieuwenhuizen:JPAMG_20:1987,Ernst:JPAMG_20:1987}.

\textit{Fluctuations of the tracer along the force.---}
We decompose the mean displacement $\langle \Delta x(t) \rangle$ contained in the variance
[Eq.~\eqref{eq:variance_general}] into the bare drift $v_0 t$ with bare velocity $v_0 = (a/2\tau)\sinh(F/2)$
for the empty lattice and a correction:
\begin{align} \label{eq:mean_displacement}
\langle \Delta x(t) \rangle = v_0 t + n \int_0^t \diff t'\Delta v(t') ,
\end{align}
where $\Delta v(t)$ is the first-order-density response for the average velocity~\cite{Leitmann:PRL_111:2013}.
Similarly, the mean-square displacement along the force 
\begin{align} 
\langle\Delta x(t)^2\rangle = 2 D_x^0 t + (v_0 t)^2 + n \Delta  R_x(t),
\end{align}
contains a diffusive contribution with bare diffusion coefficient $D_x^0 =
(a^2/4\tau)\cosh(F/2)$, drift $(v_0 t)^2$ and first-order-density response $\Delta R_x(t)$.
The bare diffusion coefficient $D_x^0$ and the bare velocity $v_0$ are connected by a Stokes-Einstein relation to linear
order in $F$. After squaring the
mean displacement [Eq.~\eqref{eq:mean_displacement}] and retaining only terms up to first order in the obstacle density,
we obtain the variance
\begin{align}
\text{Var}_x(t) = 2 D_x^0 t + n \Bigl[\Delta R_x(t) - 2v_0 t \int_0^t\diff t' \Delta v(t')\Bigr].
\end{align}
We have calculated the terms in the square bracket analytically in the frequency domain by solving for the scattering matrix in the
single obstacle case along the lines of Ref.~\cite{Leitmann:PRL_111:2013}. The new term $\Delta R_x(t)$ is essentially obtained as a
sum over certain matrix elements of the single-scattering matrix (see Supplemental Material~\cite{[{See Supplemental
Material which includes Ref.~\cite{Rossum:RMP_71:1999}, for details on the analytic solution, nonanalytic
behavior, mobile obstacles perfoming an asymmetric motion, and the asymptotic model}]Supplemental_Material}).

For increasing strength of the force, the variance shows a significant increase of the fluctuations parallel to the
force~[Fig.~\subref{fig:varx_d1e-3}{1(a)}]. In particular, at intermediate times, the fluctuations are goverened by a marked increase faster
than diffusion $\sim t$. Only at later times we recover diffusional behavior, however with a
vastly increased diffusion coefficient.

The time-dependent behavior of the variance can be quantified in more detail by considering the 
time-dependent diffusion coefficient [Fig.~\subref{fig:diffcoeffx_d1e-3}{1(c)}]
\begin{align} \label{eq:def_diffcoeffx}
D_x(t) := \frac{1}{2}\frac{\diff}{\diff t} \text{Var}_x(t),
\end{align} 
and the local exponent $\alpha\equiv \alpha(t)$ [Fig.~\subref{fig:subsuperdiffx_d1e-3}{1(b)}] defined by a logarithmic time derivative
\begin{align} \label{eq:local_exponent}
\alpha(t) := \frac{\diff \ln(\text{Var}_x(t))}{\diff\! \ln(t)} = \frac{2 D_x(t) t}{\text{Var}_x(t)} .
\end{align}
Thus, ordinary diffusion corresponds to $\alpha = 1$, whereas subdiffusive and superdiffusive behavior
is indicated by $\alpha < 1$ and $\alpha > 1$, respectively. While there is still a subdiffusive regime at small
times of the order of the density $n$, transport at strong driving is dominated by a superdiffusive regime
which grows with increasing strength of the driving [Fig.~\subref{fig:subsuperdiffx_d1e-3}{1(b)}]. 

The emergence of superdiffusion for large forces can be rationalized by observing that up to times $\tau$, the time a
tracer needs to go around an obstacle, the particle essentially moves only along the field until it hits an obstacle.
Thus, up to $\tau$ only the forward motion needs to be taken into account and the dynamics is along one-dimensional
lanes. The probability distribution then reads $\mathbb{P}(\Delta x = a\cdot j) = n(1-n)^j + (1-n)^{j+1}\delta_{jJ}$ for
$J$ jump attempts.  Then, one can work out that asymptotically the variance in continuous time is determined by the
fluctuations of the free path length and grows as 
\begin{align} \label{eq:varx_asymptotic} 
\text{Var}_x(t) = \frac{1}{3} \frac{n a^2}{64} \exp(3F/2) \frac{t^3}{\tau^3}. 
\end{align} 
This result suggests that $\alpha = 3$ is the true asymptotic
exponent of superdiffusion for the driven lattice Lorentz model~(see Supplemental
Material~\cite{Supplemental_Material}). Matching Eq.~\eqref{eq:varx_asymptotic} to the short-time diffusion ${\approx a^2 e^{F/2}
t/4\tau}$ yields as onset time of superdiffusion $\tau^*\sim \tau e^{-F/2}/\sqrt{n}$. Therefore, the window of
superdiffusion $\tau^* \lesssim t \lesssim \tau$ grows with the force.

The instantaneous response $D_x(t \to 0)$ of the time-dependent diffusion coefficient $D_x(t)$
[Eq.\eqref{eq:def_diffcoeffx}, Fig.~\subref{fig:diffcoeffx_d1e-3}{1(c)}] is determined by the first jump event only and one
readily obtains $D_x(t\to 0) = D_x^0 (1-n)$. The long-time behavior is obtained by evaluating the first-order-response 
terms $\Delta R_x(t)$ and $\Delta v(t)$ for long times leading to the asymptotic expansions
\begin{align}
\Delta R_x(t) &= \Delta R_{x,2}\frac{t^2}{2} + \Delta R_{x,1}t + \mathcal{O}(t^0), \\
\int_0^t\diff t'\Delta v(t') &= \Delta v_{1}t + \Delta v_0 + o(t^0).
\end{align}
The expressions for the coefficients are lengthy and depend only on the force and will not be shown here. 
The analytic solution fulfills the relation
$\Delta R_{x,2} = 4 v_0 \Delta v_{1}$, such that the long-time diffusion coefficient is obtained as 
\begin{align}
D_x(t\to \infty) = D_x^0 + n\Bigl[\frac{\Delta R_{x,1}}{2} - v_0\Delta v_0\Bigr] ,
\end{align}
exact in first order of the density of obstacles $n$ and for arbitrary strong driving.  Thus, in first order of the
density, the long-time behavior is always diffusive.  Yet, the long-time diffusion coefficient increases by more
than a factor of ten already at density $n = 10^{-3}$ for the large forces in Fig.~\subref{fig:diffcoeffx_d1e-3}{1(c)}.
The strong increase of the diffusion coefficient at intermediate times is a fingerprint of the superdiffusive behavior
governing the transition to the stationary state. 

For fixed density and increasing force, deviations between the analytic and the simulation results increase. This is due
to contributions higher order in the obstacle density which become more and more important for increasing force.  Such
higher-order terms arise due to scattering events of the tracer with different obstacles and are not fully included in
the first-order theory. 

It is interesting to ask how the superdiffusive behavior emerges from the equilibrium reference system for small forces. 
In equilibrium, the dynamics of a Brownian particle satisfies global detailed balance
and the approach of the diffusion coefficient to the stationary state $D^\text{eq}_x(t) - D^\text{eq}_x(t\to\infty)$ is
described by a weighted sum of relaxing exponentials~\cite{Doi:Oxford:1999,Gardiner:Springer:2009} i.e. a completely
monotone function~\cite{Feller:Wiley:1970},
\begin{align} \label{eq:completely_monotone_time}
D^{\text{eq}}_x(t) - D^{\text{eq}}_x(t\to\infty) = \int_0^\infty e^{-\gamma t} m(\diff\gamma),
\end{align}
with a non-negative measure $m(\diff\gamma)$.
In particular, in the lattice Lorentz gas in equilibrium, the approach is governed by an algebraic decay $\sim t^{-1}$
reflecting the persistent memory in the system due to repeated interaction with the obstacle
disorder~\cite{Nieuwenhuizen:PRL_57:1986}. Taking the one-sided Fourier transform $\hat{D}^\text{eq}_x(\img \omega) =
\int_0^\infty \diff t\ e^{-\img \omega t} D^\text{eq}_x(t)$, one obtains:
\begin{align}
\hat{D}^\text{eq}_x(\img\omega) - \frac{D^\text{eq}_x(t \to \infty)}{\img \omega} =
\int_0^\infty \frac{\gamma - \img\omega}{\gamma^2 + \omega^2} m(\diff\gamma) .
\end{align}

\begin{figure}[t]
\includegraphics{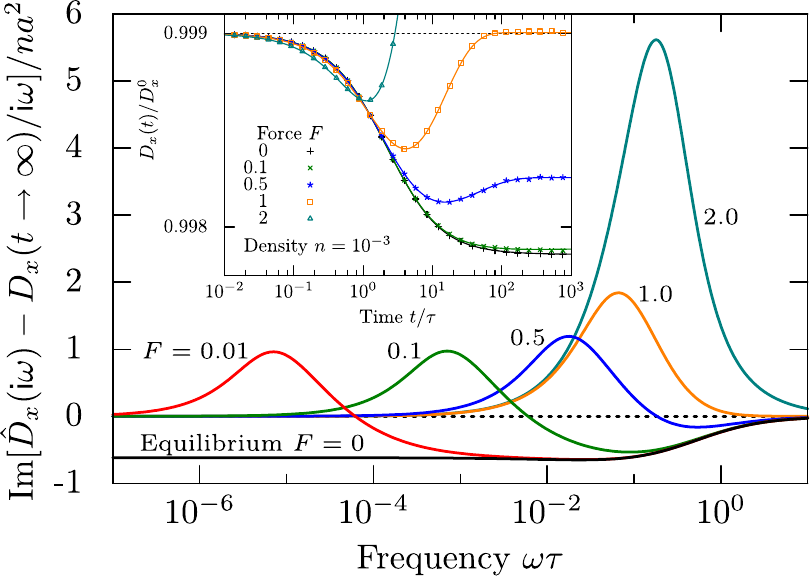}
\caption{\label{fig:freq_diffcoeffx}
Imaginary part of the frequency-dependent response encoding the approach to the stationary state for different strength
$F$ of the driving. Inset: Time-dependent diffusion coefficient for different forces exemplifying the nonmonotonic behavior.
Symbols correspond to simulation results and lines represent the theory.
}
\end{figure}

In particular, for $\omega > 0$ the imaginary part of the frequency-dependent approach to the stationary state in
equilibrium is always negative (see also Fig.~\ref{fig:freq_diffcoeffx}). The nonvanishing contribution for ${\omega\to 0}$ 
in equilibrium can be traced back to a nonanalytic small-frequency behavior of the diffusion
coefficient $\hat{D}^\text{eq}_x(\img\omega) \simeq {D_x^\text{eq}(t\to\infty)/\img\omega} -(n\pi
a^2/8)\ln(\img\omega\tau)$,
corresponding to the algebraic tail ${\sim t^{-1}}$ in the temporal domain~\cite{Nieuwenhuizen:PRL_57:1986}. Thus, for the real and imaginary part, we obtain
\begin{align}
\frac{1}{na^2}\Bigl[\hat{D}^\text{eq}_x(\img \omega) - \frac{D^\text{eq}_x(t\to\infty)}{\img \omega}\Bigr]\! \simeq
-\frac{\pi}{8}\ln (\omega\tau) - \img \frac{\pi^2}{16},
\end{align}
which rationalizes the small-frequency behavior in equilibrium~[Fig.~\ref{fig:freq_diffcoeffx}].

This behavior no longer holds true in the presence of a force on the tracer where positive contributions emerge for any
strength of the driving such that the approach to the stationary diffusion coefficient is not necessarily
monotonically decreasing~[Eq.\eqref{eq:completely_monotone_time}]. In fact, in the time domain, this deviation from equilibrium becomes
manifest in a nonmonotonic behavior of the time-dependent diffusion coefficient such that the point of least
diffusivity is always attained at intermediate times~(see inset of Fig.~\ref{fig:freq_diffcoeffx}). 

\begin{figure}[t]
\includegraphics[scale=0.99]{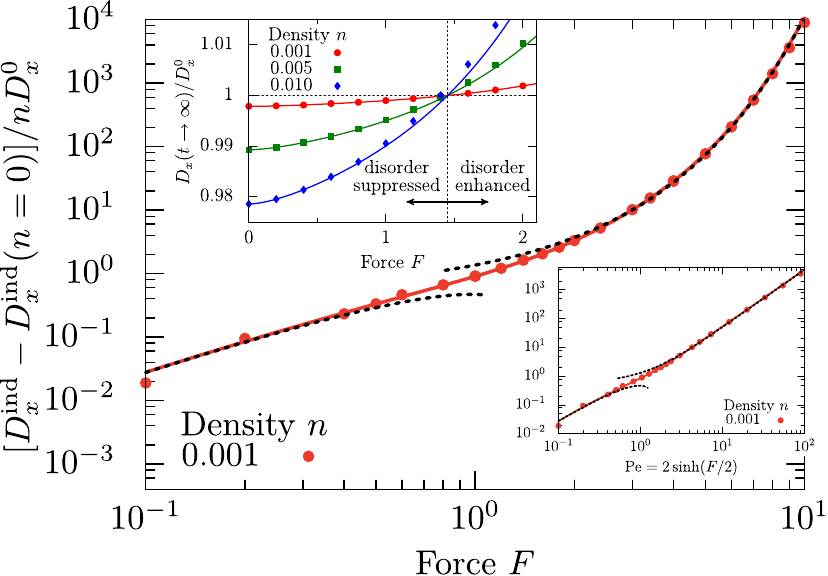}
\caption{\label{fig:stat_diffcoeffx}
Force-induced diffusion coefficient $D_x^\text{ind}$ (corrected by the empty lattice $D_x^\text{ind}(n=0)$) in units of
the bare diffusion coefficient $D_x^0$ as a function of the applied force. Symbols correspond to simulation results and
lines represent the theory. The black dashed lines represent the asymptotic expansions for small and large forces.
Inset bottom right: Same quantity as a function of the P{\'e}clet number $\text{Pe} := a v_0/D_x^0(F=0) = 2\sinh(F/2)$
given by the ratio of the bare velocity $v_0$ and the thermal fluctuations $D_x^0(F=0)$. Inset top left: Stationary diffusion
coefficient measured in units of $D_x^0$ for different forces and densities. 
}
\end{figure}

It is convenient to characterize the transport behavior in the stationary state in terms of the force-induced diffusion
coefficient $D_x^\text{ind}\equiv D_x^\text{ind}(F, n)$~\cite{Zia:JFM_658:2010} [Fig.~\ref{fig:stat_diffcoeffx}]:
\begin{align}
D_x^\text{ind} = D_x(t\to\infty) - D^\text{eq}_x(t\to\infty), 
\end{align}
with the long-time diffusion coefficient in the absence of driving, $D^\text{eq}_x(t\to\infty) =
(a^2/4\tau)\bigl[1-(\pi-1)n\bigr]$~\cite{Nieuwenhuizen:PRL_57:1986}. For small forces $F \to 0$, our explicit
solution reveals that the force-induced diffusion $D_x^\text{ind}$ acquires a leading nonanalytic term 
\begin{align}
D^\text{ind}_x = \frac{n A a^2}{4\tau}F^2[\ln(1/|F|) + B] +\mathcal{O}\bigl(F^2\ln(1/|F|)\bigr)^2,
\end{align}
with prefactor $A = (3\pi^2+4\pi+8)/16\pi \approx 0.998$  and a subleading correction term $B \equiv B(n) > 0$ (see
Supplemental Material~\cite{Supplemental_Material}). 
The origin of the nonanalytic contribution can be understood by observing that the propagators in the
presence of a force are essentially described by the equilibrium propagators up to a force-dependent shift in the frequency
domain. In particular, they inherit the nonanalytic dependence for long times from the equilibrium propagators leading
to the emergence of nonanalytic contributions for small forces in the stationary state (see
Supplemental Material~\cite{Supplemental_Material}). 

For large forces $F \gtrsim 6$, the force-induced diffusion coefficient increases
rapidly~[Fig.~\ref{fig:stat_diffcoeffx}] and assumes the asymptotic form
\begin{align}
D^\text{ind}_x = \frac{n a^2}{16\tau}\exp(3F/2) + \mathcal{O}\bigl(\exp(F)\bigr).
\end{align}
This scaling behavior can also be obtained by
asymptotic matching of the superdiffusive behavior [Eq.~\eqref{eq:varx_asymptotic}] to the diffusive increase at time scale $\tau$.

Remarkably, there exists a critical force $F_\text{c}\approx 1.45$ where in first order of the
density the long-time diffusion coefficient is identical to the bare diffusion coefficient, $D_x(t\to\infty,
F_\text{c}) = D_x^0(F_\text{c})$. Thus, this critical force separates two regimes of strikingly different
behavior also observed in other models~\cite{Zia:JFM_658:2010,Benichou:PRL_111:2013,Demery:NJP_16:2014}. For forces $F
< F_\text{c}$, the intuitive picture holds where an increase in the disorder suppresses the fluctuations, whereas in the
regime $F > F_\text{c}$
increasing disorder leads to an enhancement.

\textit{Summary and conclusion.---}
We have solved for the dynamics of a tracer particle on a lattice in response to a step force in the presence of
obstacles. The complete time dependence of the fluctuations of the tracer position parallel to the force have been
evaluated exactly for arbitrary strong driving in first order of the obstacle density. 

Our main result is the emergence of a superdiffusive growth of the fluctuations in an intermediate time regime followed
by ordinary diffusion with a stationary diffusion coefficient enhanced by orders of magnitude.  These superdiffusively
growing fluctuations have been discovered in simulations for crowded
systems~\cite{Winter:PRL_108:2012,Winter:JCP_138:2013,Schroer:PRL_110:2013}, but have also been derived analytically in
crowded lattices in confined geometries~\cite{Benichou:PRL_111:2013} and one-dimensional kinetically constrained
models~\cite{Gradenigo:PRE_93:2016}. Our results for the driven lattice Lorentz gas demonstrate that the emergence of
superdiffusion is generic and arises due to the competition of exclusion interaction and nonlinear driving already at low
obstacle densities.  In the lattice Lorentz model, the superdiffusion can be traced back to the rapid increase of the
variance of the free path lengths as the particle performs a purely directed motion along the field until it hits an
obstacle.  The full time-dependent solution additionally provides the first direct access to the intermediate window of
superdiffusive motion.

In the lattice Lorentz gas in equilibrium, global detailed balance holds and correlation functions purely relax; i.e.,
they are completely monotone. Switching on the step force drives the system out of equilibrium and the approach of the
diffusion coefficient to the stationary state is nonmonotonic, in striking contrast to the equilibrium paradigm. While
this behavior is obvious for strong forces where the stationary diffusion coefficient exceeds the initial one, an
analysis in the frequency domain reveals that this applies to arbitrarily small forces.

The frequency-dependent diffusion coefficient becomes a nonanalytic function for small frequency and for small forces.
Such singular behavior in the frequency domain has been known for transport coefficients and is related to persistent
memory effects and long-time tails~\cite{Alder:PRA_1:1970, Ernst:PLA_34:1971, vanBeijeren:RMP_54:1982,
Nieuwenhuizen:PRL_57:1986}, and more recently for the mobility also as a function of the
force~\cite{Leitmann:PRL_111:2013}. The same mechanism of repeated encounters with the same scatterer applies to the
diffusion coefficient such that the singular frequency behavior in equilibrium and the nonanalytic contribution to the
stationary diffusion coefficient are two sides of the same coin.

It is also interesting to compare the lattice Lorentz model to a dilute colloidal solution of hard spheres performing
Brownian motion. Here, the stationary diffusion coefficients have been calculated to first order in the packing fraction
for small as well as very large forces~\cite{Zia:JFM_658:2010}. Our force-induced diffusion coefficient $D_x^\text{ind}$
displays overall similar behavior, yet there the leading correction for small forces is analytic
$\mathcal{O}(\text{Pe}^2)$, where the P{\'e}clet number $\text{Pe}$ is a dimensionless measure for the force. Presumably
nonanalytic contributions arise at the next-leading order similar to the mobility
$\mathcal{O}(|\text{Pe}|^3)$~\cite{Squires:PoF_17:2005}. We anticipate that these differences originate from comparing 2D
to 3D systems~\cite{Franosch:CP_375:2010}. For large P{\'e}clet numbers, the fluctuations in the colloidal case grow
$\mathcal{O}(\text{Pe})$~\cite{Zia:JFM_658:2010}, while the dominant contribution in the lattice Lorentz system grows
much faster. However, a direct comparison of the P{\'e}clet number with our dimensionless force is dubious, since the
definition of the P{\'e}clet number involves a length scale. Furthermore, in the lattice case, local detailed balance
requires enhancing the rates exponentially.  Nevertheless, defining the P{\'e}clet number for the lattice system by
the ratio of the bare velocity $v_0$ and the thermal fluctuations $D_x^0(F=0)$, 
$\text{Pe} := a v_0/D_x^0(F=0) = 2\sinh(F/2)$, the force-induced diffusion coefficient increases much less rapidly
$\mathcal{O}(\text{Pe}^2)$ for strong driving (see inset of Fig.~\ref{fig:stat_diffcoeffx}).

While our analytic calculation is for fixed disorder, one can readily address the mobile case in simulations where we
use a mean waiting time of $\tau$ for the moving obstacles. Qualitatively, this does not affect the superdiffusive
behavior and the enhancement of the stationary diffusivity (see Fig.~\ref{fig:varx_d1e-3}), even if the jump
probabilities are asymmetric (see Supplemental Material~\cite{Supplemental_Material}). Therefore, we conclude, that the
presence of frozen disorder is not essential; rather, the effects discussed and rationalized by the theory are generic
features of the interplay of obstruction and driving.

\begin{acknowledgments}
We gratefully acknowledge support by the DFG research unit FOR1394 ``Nonlinear response to probe vitrification.''
\end{acknowledgments}


%

\onecolumngrid
\clearpage

\section{Supplemental Material}

\subsection{Analytic solution of the free dynamics}

We define the lattice with lattice spacing $a$ by the collection of all sites $\Lambda = \{\vec{r} = (ax, ay) \in
a\mathbb{Z}\times a\mathbb{Z} : x, y \in [-L/2, L/2[\}$. We employ periodic boundary conditions and anticipate the limit
of large lattices $L\to\infty$. The conditional probability $\langle\vec{r}|\hat{U}_0(t)|\vec{0}\rangle$ for a tracer
starting at the origin $\vec{0}$ and moving a distance $\vec{r}$ in lag time $t$ is determined by the time-evolution
operator $\hat{U}_0(t)$ which fulfills the master equation $\partial_t \hat{U}_0(t) = \hat{H}_0 \hat{U}_0(t)$ with
free 'Hamiltonian'
$\hat{H}_0$:
\begin{align}
\hat{H}_0 = \frac{\Gamma}{\tau}\sum_{\vec{r} \in \Lambda}\bigl[-|\vec{r}\rangle\langle\vec{r}| +
\sum_{\vec{d}\in\mathcal{N}}W(\vec{d})|\vec{r}\rangle\langle\vec{r}-\vec{d}|\bigr], \quad \mathcal{N} = \{\pm
a\vec{e}_x, \pm a\vec{e}_y\} .
\end{align}
The invariance of the free dynamics under translations becomes manifest in the plane wave basis
\begin{align}
|\vec{k}\rangle = \frac{1}{\sqrt{N}}\sum_{\vec{r}\in\Lambda}\exp(\img\vec{k}\cdot\vec{r})|\vec{r}\rangle ,
\end{align}
with number of lattice sites $N = L^2$, wave vector $\vec{k} = (k_x, k_y) \in \Lambda^* = \{(2\pi x/aL, 2\pi y/aL) : (x, y) \in \Lambda\}$, and scalar product $\vec{k}\cdot\vec{r} = k_x x + k_y y$. 
Then, the invariance of the free Hamiltonian under translations reads $\langle \vec{k} | \hat{H}_0 | \vec{k}'\rangle =
\epsilon(\vec{k})\delta(\vec{k},\vec{k}')$ with eigenvalue:
\begin{align}
\epsilon(\vec{k}) &= -\frac{\Gamma}{\tau}\sum_{\vec{d}\in\mathcal{N}} \bigl[\bigl(1-\cos(\vec{k}\cdot\vec{d}\big)\bigr) +
\img\sin(\vec{k}\cdot\vec{d})\bigr]W(\vec{d}) 
\end{align}

The moment-generating function $F_0(\vec{k},t)$ of the tracer displacements
$\Delta\vec{r} = \vec{r} - \vec{r}'$ with initial distribution ${\langle \vec{r}|p_\text{eq}\rangle = 1/N}$ is then
obtained by the matrix elements of the time-evolution operator in the plane wave basis:
\begin{align}
F_0(\vec{k},t) = \sum_{\vec{r},\vec{r}'\in\Lambda}
e^{-\img\vec{k}\cdot(\vec{r}-\vec{r}')}\langle \vec{r}|\hat{U}_0(t)|\vec{r}'\rangle\langle\vec{r}'|p_\text{eq}\rangle =
\langle\vec{k}|\hat{U}_0(t)|\vec{k}\rangle = \langle\vec{k}|\exp(\hat{H}_0 t)|\vec{k}\rangle = \exp\bigl(\epsilon(\vec{k})t\bigr),
\end{align}
where we used the formal solution $\hat{U}_0(t) = \exp(\hat{H}_0 t)$ of the time-evolution operator. In particular, we
obtain the first two moments for the displacement along the field via
\begin{align}
\langle \Delta x(t)\rangle &= -\img \frac{\partial}{\partial k_x} F_0(\vec{k},t)\Big|_{\vec{k}=0} =
\frac{a}{2\tau}\sinh(F/2)t = v_0 t, \\
\langle \Delta x^2(t)\rangle &= - \frac{\partial^2}{\partial k_x^2} F_0(\vec{k},t)\Big|_{\vec{k}=0} =
2 D_x^0 t + (v_0 t)^2, \quad D_x^0 = \frac{a^2}{4\tau}\cosh(F/2).
\end{align}

\subsection{Analytic solution to first order in the obstacle density}

In the presence of obstacles, the moment-generating function $F(\vec{k},t)$ 
of the displacements $\Delta\vec{r} = \vec{r} - \vec{r}'$ is defined in terms of the disorder-averaged
time-evolution operator $[\hat{U}(t)]_\text{av}$:
\begin{align}
F(\vec{k},t) = \sum_{\vec{r},\vec{r}'\in\Lambda} \exp[-\img\vec{k}\cdot(\vec{r}-\vec{r}')] 
\langle \vec{r} | [\hat{U}(t)]_\text{av}|\vec{r}'\rangle \langle\vec{r}'|p_\text{eq}\rangle ,
\end{align}
with initial site-occupation probability distribution $\langle\vec{r}'|p_\text{eq}\rangle = 1/N$. 
In the frequency domain, the moments are encoded in the Green function
\begin{align}
[G]_\text{av}(\vec{k}) = \frac{1}{G_0(\vec{k})^{-1} - \Sigma(\vec{k})}, 
\end{align}
with the free propagator $G_0(\vec{k})=\langle\vec{k}|\hat{G}_0|\vec{k}\rangle = \int_0^\infty \diff t\ e^{-\img\omega
t}\langle\vec{k}|\hat{U}_0(t)|\vec{k}\rangle = [\img\omega - \epsilon(\vec{k})]^{-1}$
and self-energy $\Sigma(\vec{k})$
which accounts for all possible interactions of the tracer with the obstacle disorder. In first order of the obstacle
density $n$, the self-energy can be expressed by the single-scattering $t$-matrix which represents repeated collisions
of the tracer with the same obstacle: $\Sigma(\vec{k}) = n N t(\vec{k}) + \mathcal{O}(n^2)$~\cite{Rossum:RMP_71:1999,
Nieuwenhuizen:PRL_57:1986, Leitmann:PRL_111:2013}.

The single-scattering $t$-matrix fulfills the relation
\begin{align}
\hat{t} = \hat{v} + \hat{v}\hat{G}_0\hat{t} = \hat{v} + \hat{t}\hat{G}_0\hat{v}, 
\end{align}
where $\hat{v}$ denotes the single obstacle potential which cancels transitions from and to the impurity site.
The scattering matrix is calculated in the real space basis by a $5\times 5$ matrix inversion problem $\langle
\vec{r}|\hat{t}|\vec{r}'\rangle = \langle \vec{r}|\hat{v}(1-\hat{G}_0\hat{v})^{-1}|\vec{r}'\rangle$ since the obstacle
potential has only nonvanishing contributions at the obstacle site and its four neighbors~\cite{Leitmann:PRL_111:2013}.
Then, the forward-scattering amplitude $Nt(\vec{k}) = N\langle\vec{k}|\hat{t}|\vec{k}'\rangle = \sum_{\vec{r},\vec{r}'}
\exp[-\img\vec{k}\cdot(\vec{r}-\vec{r}')]\langle\vec{r}|\hat{t}|\vec{r}'\rangle$ is derived by a transformation to the
plane wave basis and the Green function is obtained as 
\begin{align} \label{eq:green_function} 
[G]_\text{av}(\vec{k}) = G_0(\vec{k}) + n N t(\vec{k}) G_0(\vec{k})^2 + \mathcal{O}(n^2).
\end{align} 

To make connection to the stochastic simulation we correct the Green function for the fraction
$n$ of immobile random walkers starting at impurities by normalizing Eq.~\eqref{eq:green_function}
with $1/(1-n) = 1 + n + \mathcal{O}(n^2)$:
\begin{align} \label{eq:green_function_renorm} 
[G]_\text{av}(\vec{k}) = G_0(\vec{k}) + n [G_0(\vec{k}) + N t(\vec{k}) G_0(\vec{k})^2] + \mathcal{O}(n^2).
\end{align} 
Then, we calculate the mean displacement
\begin{align}
\frac{\partial}{\partial k_x}[G]_\text{av}\Bigr|_{\vec{k}=0} = \frac{\partial G_0}{\partial
k_x}\Bigr|_{\vec{k}=0}+ n \Bigl[\frac{\partial G_0}{\partial k_x} + \frac{\partial Nt}{\partial k_x}
G_0^2\Bigr]_{\vec{k}=0} + \mathcal{O}(n^2),
\end{align}
and the mean-square displacement
\begin{align}
\frac{\partial^2}{\partial k_x^2}[G]_\text{av}\Bigr|_{\vec{k}=0} = \frac{\partial^2 G_0}{\partial
k_x^2}\Bigr|_{\vec{k}=0} + n \Bigl[\frac{\partial^2 G_0}{\partial k_x^2} + \frac{\partial^2
Nt}{\partial k_x^2} G_0^2 + 4 G_0 \frac{\partial Nt}{\partial k_x}
\frac{\partial G_0}{\partial k_x}\Bigr]_{\vec{k}=0} + \mathcal{O}(n^2),
\end{align}
along the force in the frequency domain in first order of the density $n$. The derivatives after the $x$-component
of the wave vector $\vec{k}$ are obtained as sum over the matrix elements $\langle \vec{r}|t|\vec{r}'\rangle$:
\begin{align}
\frac{\partial Nt}{\partial k_x}\Bigr|_{\vec{k}=0} &= 
- \img \sum_{\vec{r},\vec{r}'} \vec{e}_x\cdot(\vec{r} - \vec{r}')\langle \vec{r} | \hat{t} |\vec{r}'\rangle, \\ 
\frac{\partial^2 Nt}{\partial k_x^2}\Bigr|_{\vec{k}=0} &= 
- \sum_{\vec{r},\vec{r}'} [\vec{e}_x\cdot(\vec{r}-\vec{r}')]^2\langle \vec{r} | \hat{t} |\vec{r}'\rangle.
\end{align}

\subsection{Nonanalytic behavior}

The conditional probability in real space can be calculated analytically~\cite{Haus:PR_150:1987} and is given by 
\begin{align} \label{eq:time_evolution_real_space}
\langle \vec{r}|\hat{U}_0(t) |\vec{0}\rangle 
= e^{F x/2a}e^{-\Gamma t/\tau} I_{x/a}(t/2\tau) I_{y/a}(t/2\tau), \quad \vec{r} = (x, y), \quad \vec{0} = (0, 0), 
\end{align}
where $I_m(\cdot)$ denotes the modified Bessel function of integer order $m$.
The free propagators $\hat{G}_0(\img\omega)$ are defined by a one-sided Fourier transform
$\langle \vec{r} | \hat{G}_0(\img\omega) | \vec{0}\rangle = \int_0^\infty \diff t\ e^{-\img\omega t} \langle \vec{r} |
\hat{U}_0(t) | \vec{0}\rangle$
and encode the time-evolution of the system in the frequency domain. Then, one observes that the propagator in the case
of driving is essentially obtained by the equilibrium propagator $\hat{G}_0^\text{eq} = \hat{G}_0(F=0)$ via
\begin{align}
\langle \vec{r} | \hat{G}_0(\img\omega) | \vec{0}\rangle 
= e^{Fx/2a} \int_0^\infty \diff t\ e^{-[\img\omega + (\Gamma - 1)/\tau] t} \langle \vec{r}|\hat{U}_0(t,F=0)|\vec{0}\rangle  
= e^{Fx/2a} \langle \vec{r} | \hat{G}^\text{eq}_0(\img\Omega) | \vec{0}\rangle .
\end{align}
The only difference is a site-dependent prefactor and a shift in the frequency of the equilibrium propagator
$\img\Omega = \img\omega + (\Gamma - 1)/\tau$. For example, the propagator 
$\langle\vec{0}|\hat{G}_0^\text{eq}(\img\omega)|\vec{0}\rangle = 
(2\tau/\pi) \text{K}[(1+\img\omega\tau)^{-2}]/(1+\img\omega\tau)$
can be expressed by the complete elliptic integral of the first kind $\text{K}[x]= \int_0^{\pi/2} \diff \theta\ [1 -
x\sin^2(\theta)]^{-1/2}$ and has the following nonanalytic expansion for $\img\omega \to 0$ and $-\pi/2 <
\text{arg}(\img\omega) < \pi/2$:
\begin{align}
\langle\vec{0}|\hat{G}_0^\text{eq}(\img\omega)|\vec{0}\rangle 
= \frac{\tau}{\pi}\ln(8/\img\omega\tau) +
\frac{\tau}{2\pi}\img\omega\tau[1-\ln(8/\img\omega\tau)] + \mathcal{O}\bigl(\omega^2\ln(\omega)\bigr). 
\end{align}
Thus, for long times $\img\omega \to 0$, and small forces $\Gamma - 1 = F^2/16 + F^4/768 + \mathcal{O}(F^6)$, the
propagators in the presence of a force inherit the nonanalytic dependence from the propagators in equilibrium:
\begin{align}
\langle \vec{0} | \hat{G}_0(\img\omega) | \vec{0}\rangle 
= \frac{\tau}{\pi}\ln(128/F^2) +
\frac{\tau}{2\pi}\frac{F^2}{48}[1-3\ln(128/F^2)] + \mathcal{O}\bigl(F^4\ln(1/F)\bigr),\quad F\downarrow 0. 
\end{align}
Since the stationary diffusion coefficient essentially results from solving a
$5\times 5$ matrix problem with the free propagators as entries, the stationary
diffusion coefficient displays nonanalytic contributions of the same type:
\begin{align}
D_x(t\to\infty) = D_x^\text{eq}(t\to\infty) + \frac{n A a^2}{4\tau}F^2[\ln(1/|F|) + B]
+\mathcal{O}\bigl(F^2\ln(1/|F|)\bigr)^2, 
\end{align}
with equilibrium diffusion coefficient $D_x^\text{eq}(t\to\infty) = (a^2/4\tau)[1-(\pi-1)n]$. The subleading correction $B\equiv B(n)$ can be explicitly evaluated to
\begin{align}
B = \frac{7}{2}\ln(2) - \frac{4\pi^4 - \pi^3 - 6\pi^2 - 24\pi}{2(\pi-2)(3\pi^2+4\pi+8)} + \frac{2\pi/n}{3\pi^2 + 4\pi + 8}.
\end{align}

\subsection{Asymmetric simple exclusion process for mobile obstacles}

\begin{figure}[h]
\includegraphics[scale=1.0]{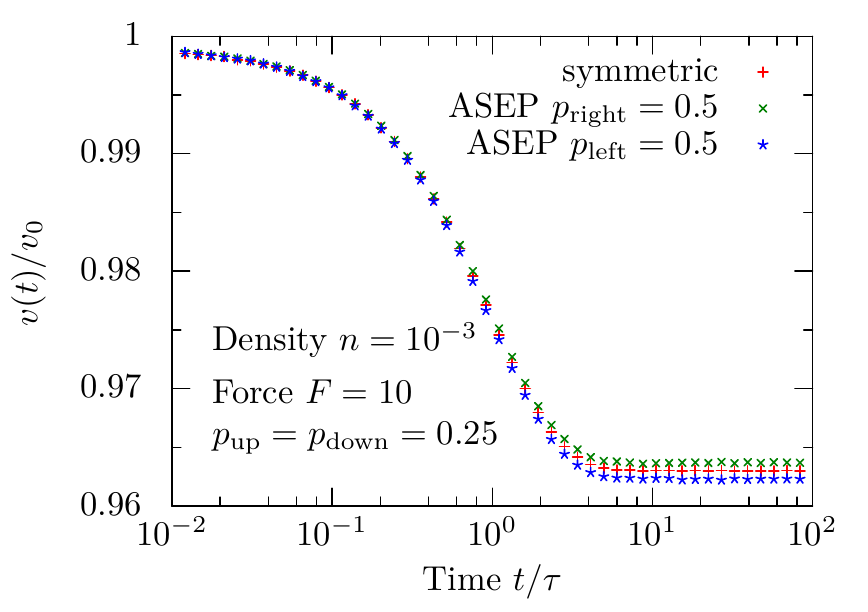}
\includegraphics[scale=1.0]{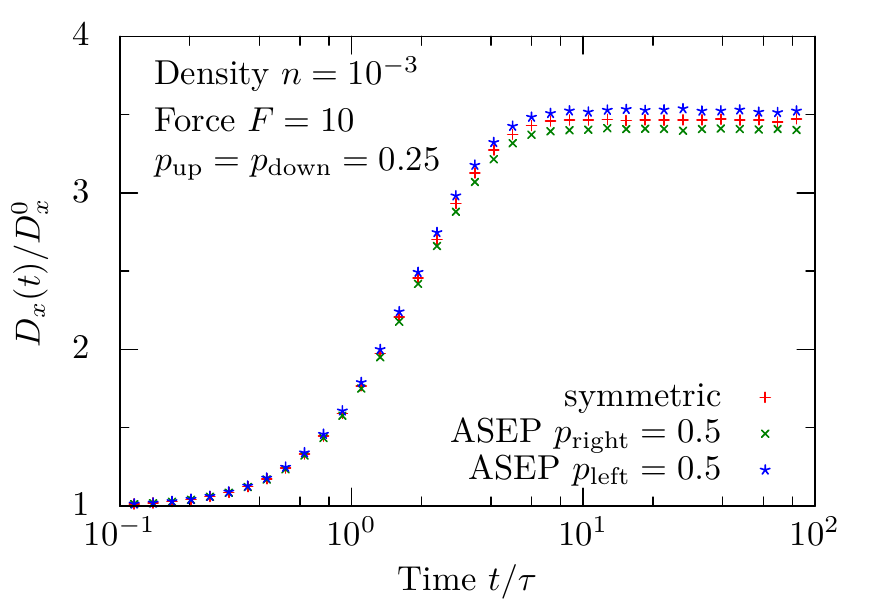}
\caption{\label{FigAsep}
Time-dependent velocity $v(t)$ and diffusion coefficient $D_x(t)$ for the case that the obstacles perform an asymmetric simple exclusion process. The jump probability
perpendicular to the field is always symmetric and given by $p_\text{up}=p_\text{down}=1/4$. The curves shown correspond to $p_\text{right}=p_\text{left}=1/4$ (symmetric),
bias in direction of the field $p_\text{right}=1/2$, $p_\text{left}=0$, and bias in the opposite direction $p_\text{left}=1/2$, $p_\text{right}=0$.
}
\end{figure}

We have simulated the dynamics of the tracer in the presence of mobile obstacles performing an asymmetric simple exclusion process~[Fig.~\ref{FigAsep}].
The velocity and the diffusion coefficient behave rather similarly to the case of unbiased mobile obstacles.
In particular, the effect of giant diffusion accompanied by a crossover regime persists also for the asymmetric case.

\subsection{Asymptotic model}

For large forces, the transition rate along field dominates the transport behavior and the motion of the tracer
perpendicular and against the field can be ignored. Hence, in every jump the tracer hits an obstacle with probability $p
= n$ and the probability for a displacement $\Delta x$ after $J$ jumps is given by
\begin{align}
\mathbb{P}(\Delta x = a\cdot j) = q^j p + q^j(1-p)\delta_{jJ} = p + \delta_{jJ}[1-(J+1)p] + \mathcal{O}(p^2),\quad j = 0, \dotsc, J,\quad q = 1-p. 
\end{align}
Then, one can readily calculate the mean and mean-square displacement of the tracer:
\begin{align}
\langle \Delta x_J\rangle &= \sum_{j=0}^J aj \mathbb{P}(\Delta x = a\cdot j) = \frac{1}{2}na J(J+1) + aJ[1-(J+1)n] + \mathcal{O}(n^2), \\
\langle \Delta x^2_J \rangle &= \sum_{j=0}^J (aj)^2 \mathbb{P}(\Delta x = a\cdot j) = \frac{1}{6} na^2 J(J+1)(2J+1) +
a^2J^2[1-(J+1)n] + \mathcal{O}(n^2) .
\end{align}
After performing the transformation to continuous time via the Poisson transform with a mean waiting time of $\tau/\gamma = 4\tau/e^{F/2}$, 
\begin{align}
\langle \Delta x (t)\rangle^2 &= 
\biggl[\sum_{J=0}^\infty \langle \Delta x_J \rangle \frac{(\gamma t/\tau)^J}{J!} e^{-\gamma t/\tau}\biggr]^2 =
-na^2\Bigl(\frac{\gamma t}{\tau}\Bigr)^3 - 2 na^2 \Bigl(\frac{\gamma t}{\tau}\Bigr)^2 + a^2\Bigl(\frac{\gamma
t}{\tau}\Bigr)^2 + \mathcal{O}(n^2), \\
\langle \Delta x^2 (t)\rangle &= \sum_{J=0}^\infty \langle \Delta x^2_J \rangle \frac{(\gamma t/\tau)^J}{J!} e^{-\gamma t/\tau} = -\frac{2}{3}na^2\Bigl(\frac{\gamma t}{\tau}\Bigr)^3 -\frac{5}{2}na^2\Bigl(\frac{\gamma t}{\tau}\Bigr)^2 +
a^2(1-n)\Bigl(\frac{\gamma t}{\tau}\Bigr) + a^2\Bigl(\frac{\gamma t}{\tau}\Bigr)^2 
+ \mathcal{O}(n^2), 
\end{align}
we obtain the variance of the displacements in first order of the density $n$:
\begin{align} \label{eq:AsymSol}
\langle \Delta x^2(t) \rangle - \langle \Delta x(t) \rangle^2 = \frac{1}{3}\frac{n a^2}{64} \exp(3F/2) \frac{t^3}{\tau^3} - 
\frac{1}{2}\frac{n a^2}{16} \exp(F)\frac{t^2}{\tau^2} + \frac{a^2}{4}(1-n) \exp(F/2) \frac{t}{\tau}+ \mathcal{O}(n^2) .
\end{align}

\begin{figure}[h]
\includegraphics{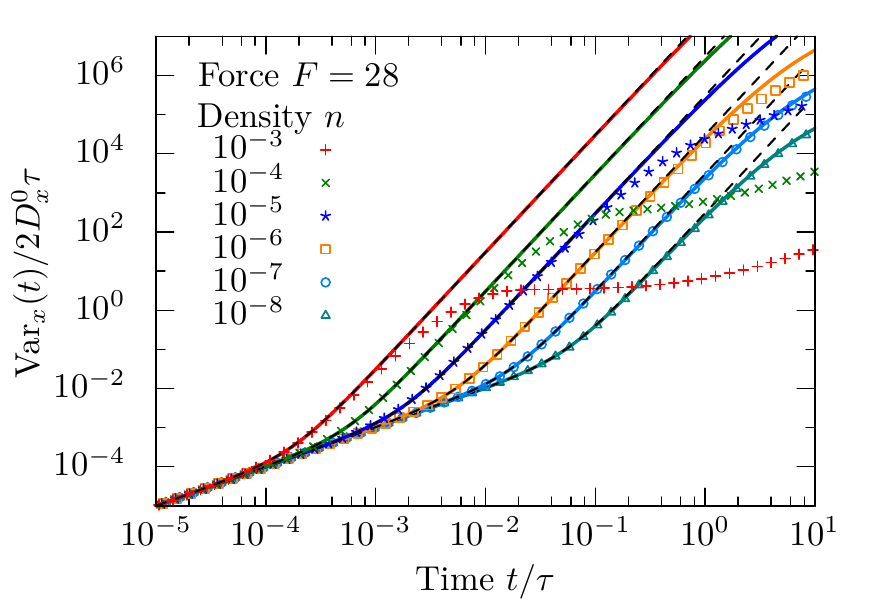}
\includegraphics{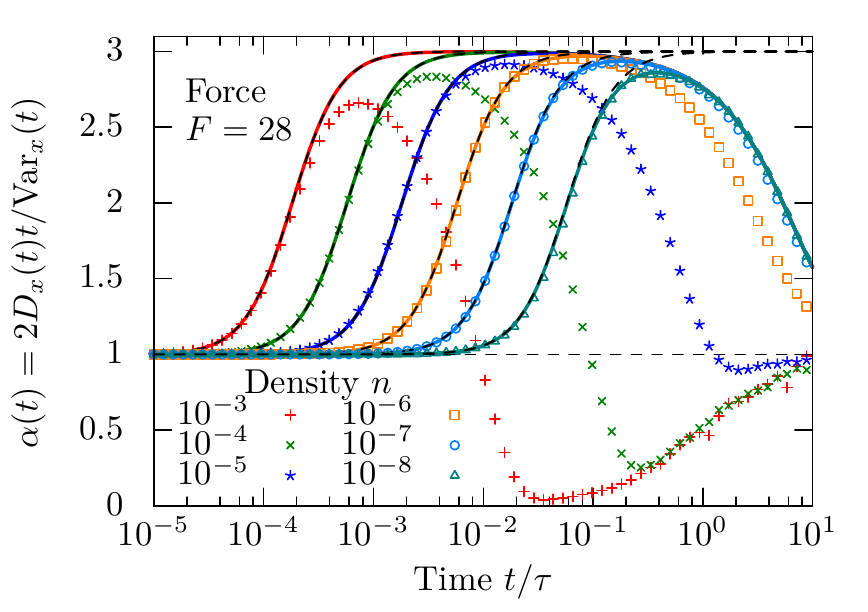}
\caption{\label{FigCompSimAsym}
Time-dependent variance $\text{Var}_x(t)$ and local exponent $\alpha(t)$ for fixed force $F=28$ and different densities. 
Solid lines correspond to the full time-dependent solution in first order and symbols represent simulation
results. The black dashed lines correspond to the asymptotic model~[Eq.~\eqref{eq:AsymSol}]. Density decreases from left to right.
}
\end{figure}

For large forces, the asymptotic model captures the dynamics of the first-order solution quantitatively until the diffusive
motion perpendicular to the force becomes relevant $t\gtrsim\tau$~[Fig.~\ref{FigCompSimAsym}]. Hence, for intermediate times,
the dynamics becomes asymptotically superdiffusive $\sim t^\alpha$ with exponent $\alpha = 3$. We also performed
simulations at high forces for different densities where the exponent $\alpha = 3$ can be observed in
simulations.

\end{document}